\documentclass[journal]{IEEEtran}
\ifCLASSINFOpdf
\else
\fi
\usepackage{amsmath}
\usepackage{algorithmic}
\usepackage{url}
\usepackage{array}

\usepackage{hyperref} 
\hypersetup{
    colorlinks=true,
    linkcolor=blue,
    filecolor=magenta,      
    urlcolor=blue,
}
\usepackage[utf8x]{inputenc}
\usepackage{booktabs} 
\usepackage{eurosym} 
\usepackage{csquotes} 
\usepackage{import} 
\usepackage{cite}
\usepackage{bm}
\usepackage{amssymb}

\usepackage{dirtytalk}

\usepackage{cleveref}

\usepackage{multirow}
\usepackage{float} 

\usepackage[pdftex]{graphicx}
\graphicspath{{./Figures/}}
\DeclareGraphicsExtensions{.pdf,.jpeg,.png}

\usepackage[caption=false,font=normalsize,labelfont=sf,textfont=sf]{subfig}
\usepackage{color,soul}

\usepackage{bbm}

\usepackage{bbold}

\usepackage{longtable}

\newcommand{\abs}[1]{\lvert #1 \rvert}

\usepackage{derivative}

\usepackage[pdftex,dvipsnames]{xcolor}

\usepackage{xargs}
\newcommandx{\todelete}[2][1=]{\todo[linecolor=red,backgroundcolor=red!,bordercolor=red,#1]{#2}}
\newcommandx{\torethink}[2][1=]{\todo[linecolor=orange,backgroundcolor=orange!,bordercolor=orange,#1]{#2}}
\newcommandx{\tochange}[2][1=]{\todo[linecolor=cyan,backgroundcolor=cyan!,bordercolor=cyan,#1]{#2}}
\newcommandx{\tocheck}[2][1=]{\todo[linecolor=yellow,backgroundcolor=yellow!,bordercolor=yellow,#1]{#2}}
\newcommandx{\toimprove}[2][1=]{\todo[linecolor=Plum,backgroundcolor=Plum!,bordercolor=Plum,#1]{#2}}
\newcommandx{\thiswillnotshow}[2][1=]{\todo[disable,#1]{#2}}

\usepackage{mathtools}

\usepackage[abbreviations,symbols,nomain,toc,postdot,stylemods]{glossaries-extra}

\GlsXtrEnableEntryCounting
 {abbreviation}
 {1}

\newabbreviation{apd}{APD}{active power disturbance}
\newabbreviation{aps}{APS}{autonomous power system}

\newabbreviation{bps}{BPS}{bulk power system}

\newabbreviation{capex}{CAPEX}{capital expenditure}
\newabbreviation{cdf}{CDF}{cumulative distribution function}
\newabbreviation{coi}{COI}{center of inertia}

\newabbreviation{ed}{ED}{economic dispatch}
\newabbreviation{ems}{EMS}{energy management system}
\newabbreviation{ess}{ESS}{energy storage system}

\newabbreviation{fcuc}{FCUC}{frequency-constrained unit commitment}
\newabbreviation{fcr}{FCR}{frequency containment reserves}
\newabbreviation{frr}{FRR}{frequency restoration reserves}

\newabbreviation{ghg}{GHG}{greenhouse gases}
\newabbreviation{gt}{GT}{gas turbine}

\newabbreviation{milp}{MILP}{mixed integer linear programming}
\newabbreviation{minlp}{MINLP}{mixed integer nonlinear problem}

\newabbreviation{og}{O\&G}{oil and gas}
\newabbreviation{opex}{OPEX}{operational expenditure}

\newabbreviation{res}{RES}{renewable energy source}
\newabbreviation{rocof}{RoCoF}{rate of change of frequency}

\newabbreviation{soe}{SoE}{state of energy}
\newabbreviation{soc}{SoC}{state of charge}

\newabbreviation{sos}{SOS}{special ordered set}

\newabbreviation{pfc}{PFC}{primary frequency control}

\newabbreviation{uc}{UC}{unit commitment}

\begin{document}
%
\title{Optimal Energy Management in Autonomous Power Systems with Probabilistic Security Constraints and Adaptive Frequency Control}

%
%
%

\author{Spyridon~Chapaloglou,
        Erick~Alves,~\IEEEmembership{Senior Member,~IEEE},        Vincenzo~Trovato,~\IEEEmembership{Member,~IEEE},
        and~Elisabetta~Tedeschi,~\IEEEmembership{Senior~Member,~IEEE}
\thanks{This research was funded by: 1) VISTA - a basic research program in collaboration between The Norwegian Academy of Science and Letters, and Equinor; 2) the Onassis Foundation - Scholarship ID: F ZP 056-1/2019-2020; 3) the Research Council of Norway under the program PETROMAKS2, grant number 281986, project ``Innovative Hybrid Energy System for Stable Power and Heat Supply in Offshore Oil and Gas Installation (HESOFF)''.}%
\thanks{S. Chapaloglou, E. Alves, and E. Tedeschi are with the Department of Electric Power Engineering, Norwegian University of Science and Technology, 7034 Trondheim, Norway, e-mail: (see \url{https://www.ntnu.edu/employees/spyridon.chapaloglou}). V. Trovato is with the Department of Civil, Environmental and Mechanical Engineering at University of Trento and with the Department of Electrical and Electronic Engineering at Imperial College London. E. Tedeschi is with the Department of Industrial Engineering, University of Trento, 38123 Trento, Italy.}}

%
%

\markboth{Journal of \LaTeX\ Class Files,~Vol.~14, No.~8, August~2015}%
{Shell \MakeLowercase{\textit{et al.}}: Bare Demo of IEEEtran.cls for IEEE Journals}
%



\maketitle
\begin{abstract}
    The decarbonization of many heavy power-consuming industries is dependent on the integration of renewable energy sources and energy storage systems in isolated autonomous power systems.
The optimal energy management in such schemes becomes harder due to the increased complexity and stability requirements, the rapidly varying operating conditions and uncertainty of renewable sources, the conflicting objectives across different timescales, the limited amount of reliable power sources and energy storage.
The state of charge management when energy storage is used for multiple services, such as optimal scheduling and frequency support, is one of the most notorious problems in this context.
To address this issue, an optimal energy management system is proposed in this paper.
It co-optimizes the primary frequency control layer and the dispatch schedule of conventional generators and energy storage by taking advantage of an algorithm that provides adaptive active power demand uncertainty quantification, theoretical guarantees for frequency stability, and bounds for the reserves for frequency support assigned to the energy storage system.
A convex reformulation is derived enabling the efficient solution of the involved optimization problem, being a test case of an isolated offshore oil and gas platform presented for validation.
\end{abstract}


%
\IEEEpeerreviewmaketitle

\section{Introduction}
\label{sec:Intro}

\IEEEPARstart{T}{aking} optimal decisions to reduce \textit{i)} fuel consumption,  \textit{ii)} \gls{ghg} emissions, \textit{iii)} equipment degradation and \textit{iv)} system insecurity in power systems with a high share of \glspl{res} is an intricate task.
This endeavor requires the solution of problems such as the \gls{uc}~\cite{hongUncertaintyUnitCommitment2021}, \gls{ed}~\cite{xiaOptimalDynamicEconomic2010}, and allocation of reserves for frequency control~\cite{mohandesReviewPowerSystem2019}, resulting in complex non-convex optimization formulations.
Those have typically conflicting objectives, continuous and binary decision variables, and high uncertainty from particular nodes, such as loads and \glspl{res}.
Reliability can, for instance, be increased where more dispatchable units, such as \glspl{gt}, are kept online for longer periods, as they respond well to fast variations of loads and \glspl{res}.
This action may, however, increase fuel consumption, emissions, and equipment degradation, affecting negatively \gls{opex}.

When compared to traditional \glspl{bps}, this type of optimization can have different characteristics in \glspl{aps}, such as isolated industrial plants, \gls{og} platforms, ships, islands, and community microgrids.
Due to the size and complexity of the problem, several challenges exist in \glspl{bps} to implement real-time, advanced algorithms for optimal dispatch and real-time allocation of frequency reserves~\cite{dorflerDistributedControlOptimization2019}.
Such problems can however be addressed in many \glspl{aps} due to the limited number of dispatchable power sources, which typically provide simultaneously several ancillary services.
The methods for allocation of frequency reserves and tuning of the system damping may also differ considerably between \gls{bps} and \gls{aps}.
The system damping is, in the former, adapted using binary decision variables which switch on and off units having a fixed active power-frequency droop, as usually the individual contribution of each unit to the total damping is small.
In the latter, the droop of individual units may represent a large portion of the system damping, being therefore necessary to readjust them in real-time using integer variables to obtain optimal results.

Optimization objectives and decisions in \glspl{aps} may also be coupled and affected by constraints in different time scales, being \glspl{ess} remarkable examples of equipment that enhances such dependencies.
\Glspl{ess} can, for instance, provide increased flexibility towards the optimal scheduling of fossil-based energy sources and avoid prolonged operation in partial loads, where emissions are much higher. 
They can also be assigned as spinning reserves for frequency control, which would require less \glspl{gt} on for the same system security requirements and increase environmental gains even further.
Where these two grid services are provided simultaneously, the scheduled trajectory of the \gls{ess}' \gls{soe} may be disturbed, affecting the optimality or even the feasibility of the original schedule.
To decide in \glspl{bps} the effect that the provision of frequency reserves has on the \gls{ess}' optimal schedule, scenarios such as the worst-case \gls{apd} or the N-1 criterion are many times used~\cite{comissionregulationeuGuidelineElectricityTransmission2017,mohandesReviewPowerSystem2019}.
These criteria can however be over-conservative in \glspl{aps}~\cite{polleuxAllocationSpinningReserves2022a} and the \gls{capex} necessary for a fully-fledged \gls{ess} may not justify its benefits, being some probability of load shedding and generation curtailment acceptable many times~\cite{sigristReviewStateArt2018}.
When it comes to energy management in \glspl{aps} with high penetration of \gls{res}, there is therefore a need and potential for better assessment of frequency stability requirements as well as coordination and scheduling of reserves.

\subsection{Literature review}
The integration of frequency stability constraints in the scheduling phase of \glspl{bps} with high penetration of \gls{res} has been an active area of research in recent years and is well described in the literature \cite{luo_stability-constrained_2020}.
The use of linear constraints had been a typical approach to increase the fidelity of the frequency response in the \gls{uc} and \gls{ed} problems, which are obtained by linearizations and/or analytical solutions of the swing equation model \cite{prakash_frequency_2018,shi_analytical_2018,badesa_simultaneous_2019,trovatoUnitCommitmentInertiaDependent2019,zhang_modeling_2020,oskouee_primary_2021,ding_two-stage_2021,trovatoSystemSchedulingOptimal2022}.
These procedures provide good approximations in \glspl{bps}, where the largest \gls{apd} is a small fraction of the installed capacity and transient and steady-state frequency deviations are usually required to be below 2\% of its rated value \cite{comissionregulationeuGuidelineElectricityTransmission2017}.
Frequency deviations may, on the other hand, be much higher in \gls{aps} due to low inertia and limited amount of frequency reserves when compared to the worst-case \glspl{apd}, causing the effects of non-linear dynamics to be sizeable during large disturbances \cite{alves_sufficient_2021}.
Where these non-linear effects are considered, bilinear terms are introduced in the frequency stability constraints, those requiring specialized reformulation-linearization techniques \cite{ahmadi_security-constrained_2014,wen_frequency_2016,zhang_frequency-constrained_2021}.
This increase of the optimization model complexity can play an important role in \gls{aps}, while it may bring only negligible frequency stability improvements in \gls{bps}.  

The challenge of leveraging optimal system operation and security in \glspl{aps} was also addressed recently, being alternatives presented to the deterministic evaluation of the worst-case \gls{apd} or the N-1 criterion. 
References \cite{cardozo_frequency_2017,javadi_look_2019,zhang_modeling_2020, yin_frequency-constrained_2021,safari_stochastic_2021}, for instance, considered the effect of anticipated net load variations and applied dynamic constraints for sizing \gls{fcr} and required inertia.
These works did not evaluate however the impact that re-adjusting the droop of online units, instead of switching on and off units with fixed droops, would have on frequency stability and \gls{opex}.
References \cite{prakash_frequency_2018,carrion_impact_2020,ding_two-stage_2021,zhang_frequency-constrained_2021} tried moreover to tackle the problem by applying simplified and static uncertainty models, such as non adaptive and arbitrary uncertainty intervals, distributions, and scenario selection, which have the limitation of not providing probabilistic guarantees.


\subsection{Paper contributions}
The main contribution of this paper is an algorithm for the \gls{ems} of a generalized \gls{aps} equipped with an \gls{ess} that is capable of simultaneously achieving optimal scheduling and securing system operation under dynamic uncertainty considerations and bounded impact on its optimal scheduling.
The proposed algorithm introduces realistic frequency constraints with general applicability, those not being limited to small scale power variations, combined with a probabilistic security analysis based on a novel adaptive power variation uncertainty quantification scheme.
This new method allows an \gls{ems} to allocate time-varying optimal frequency control reserves with bounded divergence from its optimal \gls{soe}.
A simple strong \gls{milp} reformulation is derived to efficiently implement the proposed algorithm, whose effectiveness is verified by simulations using the case study of a wind-powered offshore \gls{og} platform.

The rest of the paper is organized as follows. The proposed \gls{ems} algorithm and its numerical implementation are presented in \cref{sec:method}, while simulations in Matlab/Simulink are used to validate and discuss the various features of the proposed algorithm using the case study of a wind-powered offshore \gls{og} platform in \cref{sec:simulations}. The main conclusions are finally presented in \cref{sec:Conclusions}.

\section{Method}
\label{sec:method}

\begin{figure}[htb]
    \centering
    \centerline{\includegraphics[width=\columnwidth]{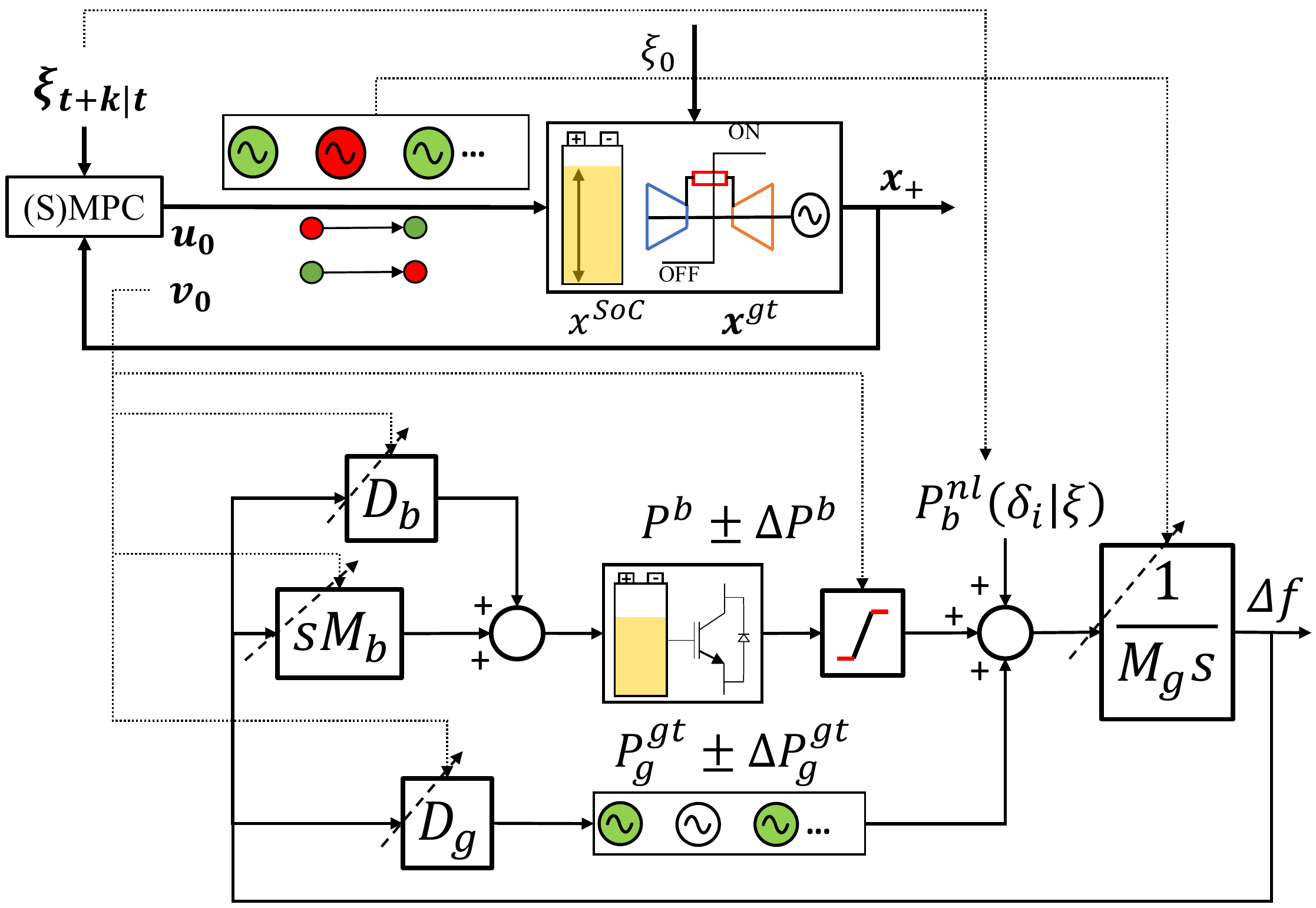}}
    \caption{Hierarchical control system schematic where the upper layer optimal discrete time control is integrated with the lower time scale continuous adaptive primary frequency control.}
     \label{fig:dstrb_control_schematics}
\end{figure}

\begin{figure}[htb]
    \centering
    \centerline{\includegraphics[width=0.8\columnwidth]{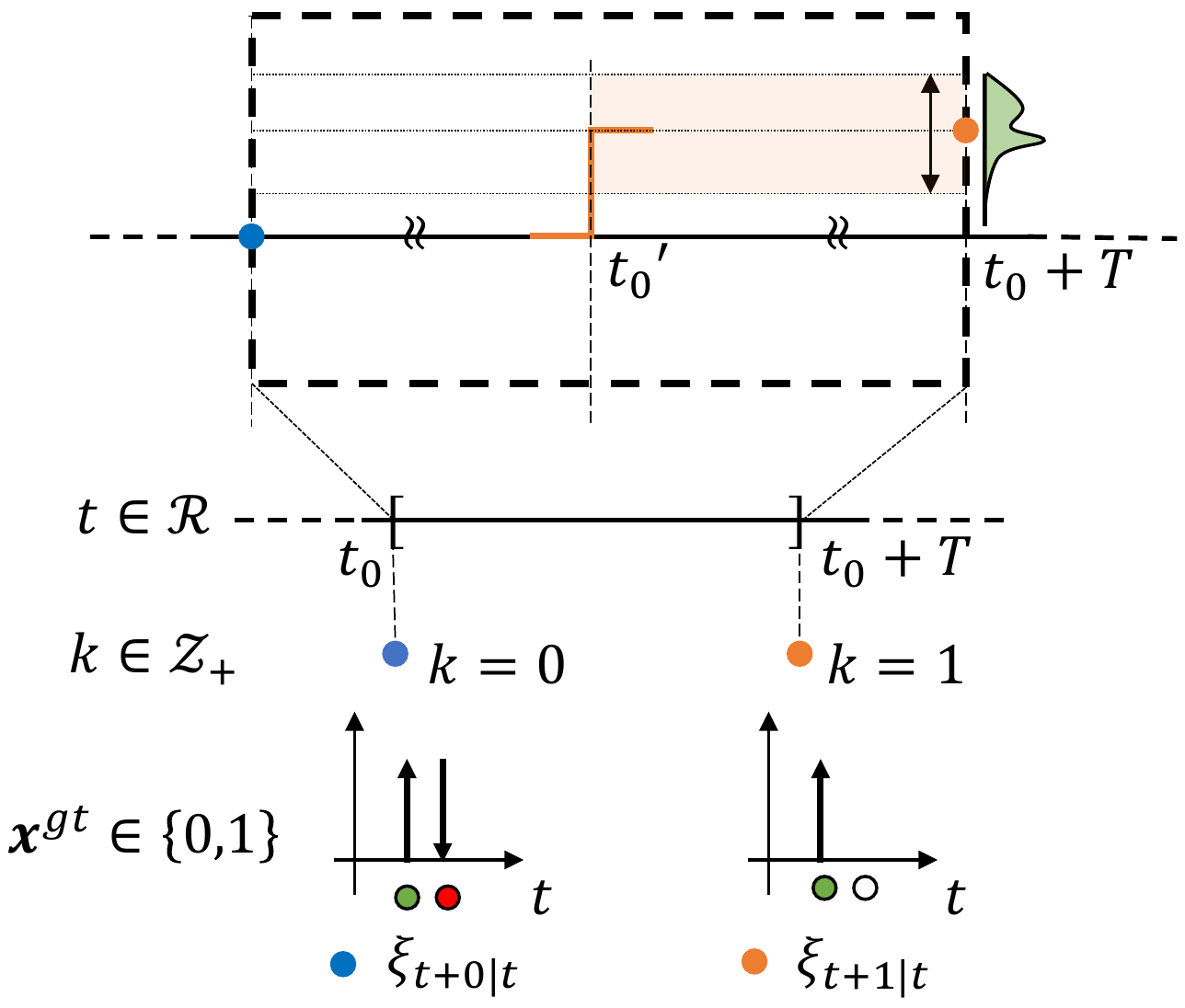}}
    \caption{Visualization of the proposed adaptive and probabilistic step-like net \glsxtrlong{apd} quantification at a time instant in the continuous range between the discrete points in time where decisions are taken.}
     \label{fig:concept}
\end{figure}

The concept and the methodology proposed in this paper is presented in detail in this section.
It assumes the existence of a centralized \gls{ems} in an \gls{aps} that is capable not only to dispatch a set of $g$ generators and a single \gls{ess} $b$, but also to decide for the primary frequency controllers the proportional gain (droop) $D_g$ of each generator, the proportional gain (droop) $D_b$ and derivative gain (virtual inertia) of the \gls{ess}, as summarized in \cref{fig:concept,fig:dstrb_control_schematics}.
To take these decisions, an adaptive and probabilistic step-like algorithm quantify net \gls{apd} at discrete time intervals $T$ is applied, as shown in \cref{fig:concept}. 

\subsection{Frequency Control and Reserves Allocation}
\label{subsec:pfc}

The non-linear dynamics of the frequency deviation for the \gls{coi} in an ac power system are given in \cref{eq:nlSwing}, where $\mathcal{X}=\frac{\omega}{\omega_s}$, and $\omega, \omega_s$ are the \gls{coi} frequency and its rated value respectively.
\begin{align}
     \dot{\mathcal{X}} &= \frac{\mathcal{D}}{\mathcal{M}} \left( \mathcal{X}-1 \right) + \frac{P^{nl}_b}{\mathcal{X}\mathcal{M}}
        \label{eq:nlSwing} \\
     \mathcal{D} &\ge \frac{P^{nl}_b}{r_{ss}(1-r_{tr})}
        \label{eq:Dmin} \\
        \mathcal{M} &\ge \frac{P^{nl}_b}{\Bar{\gamma}}   
    \label{eq:Mmin}
\end{align}

It is shown in \cite{alves_sufficient_2021} that if the total system damping $\mathcal{D}$ is selected as in \cref{eq:Dmin}, then $\mathcal{X}$ will be bounded post-disturbance by $r_{ss}$ for an instantaneous net active power imbalance $P^{nl}_b$, where $r_{tr}$ is a pre-defined constant bound for the nadir/zenith of $\mathcal{X}$ during the transient period.
Note that the step-response of frequency for active power disturbances is, in most ac power systems, underdamped due to delays of actuators and their non-linearities~\cite{ieeetaskforceonturbine-governormodelingDynamicModelsTurbineGovernors2013}.
Those effects are not modelled in \cref{eq:nlSwing} because they influence only the frequency nadir or zenith after the disturbance $P^{nl}_b$ occurs, being the post-disturbance steady-state value of $\mathcal{X}$  unaffected~\cite{alves_sufficient_2021}.  
The variables $r_{ss}$ and $r_{tr}$ in \cref{eq:Dmin} represent, in other words, the allowed steady-state and transient frequency deviations in per unit of $\mathcal{X}$, being usually defined in grid codes or by the system operator.

It is typical to assume $r_{tr} > r_{ss}$ to provide a safety margin which avoids triggering protection schemes such as under frequency load shedding or over frequency generation curtailment during large disturbances. 
When $r_{ss}=r_{tr}$, there exists no safety margin between steady-state and transient values.
It shall also be noted that \cref{eq:Dmin} is valid for any value of $\mathcal{X}$, and not only for small deviations around the operating point.

The minimum required inertia $\mathcal{M}$ is retrieved from \cref{eq:nlSwing}, assuming that a)~the disturbance $P^{nl}_b$ occurs at $t=t_0'$, b)~the system was previously in balance ($\mathcal{X} \rvert_{t=t_0'}=1$), and c)~the maximum \gls{rocof} is $\dot{\mathcal{X}} \rvert_{t=t_0'}= \Bar{\gamma}$, which results in \cref{eq:Mmin}.

Ensuring that enough power reserves are available in an \gls{aps} is a necessary condition to satisfy the constraints for frequency control imposed by \cref{eq:Dmin,eq:Mmin}.
The values of $\mathcal{D}$ and $\mathcal{M}$ depend on the number of online units and their individual parameters, being expressed by \cref{eq:Dsum,eq:Msum}, where $D_g$ and $M_g$ denote the droop setting and the inertia of each online generator, and $D_b$ and $M_b$ are the virtual damping and inertia emulated by the \gls{ess}, respectively.
The \gls{ems} must guarantee that each individual equipment has enough available power capacity to provide the frequency reserves assigned to it, what leads to the constraints in \cref{eq:Pblim,eq:Pglim}.

\begin{align}
    \mathcal{D} &= 
      \displaystyle \sum_{g}^{N_g}x_{g}^{gt} D_{g} + D_b 
        \label{eq:Dsum} \\
    \mathcal{M}  &= 
      \displaystyle \sum_{g}^{N_g}x_{g}^{gt} M_{g} + M_b
        \label{eq:Msum} \\
    \abs{\Delta P_b(t)} &\le M_b \Bar{\gamma} + D_b r_{tr}
        \label{eq:Pblim} \\
    \abs{\Delta P_g(t)} &\le x_{g}^{gt} D_g r_{tr}, \; \forall \; g \in \mathcal{N}_g
        \label{eq:Pglim}
\end{align}

\subsection{Optimal and Bounded Energy Management under Uncertainty}
\label{subsec:optProb}

\subsubsection{Allocation of frequency reserves with bounded energy storage}
\label{subsubsec:ess_bound}
The \gls{ems} must also ensure that enough energy capacity is available in the \gls{ess}.
This is required not to violate the storage limits when providing frequency control in the current scheduling step $t_0$ while charge or discharge events had already been scheduled for the next step $t_0 + T$, where $T$ is the time step of the \gls{ems}.
To deal with this situation, the constraint in \cref{eq:max_energy_dev} is proposed, where $\Delta E^{b}(t)$ is a deviation around the optimal scheduled \gls{soe} $E^{b}(t)$, and  $\lambda$ is a user-defined (absolute) percentage value.
The proposed constraint allows, in other words, an allocation of frequency reserves to the \gls{ess} that is proportional to its current energy level.

\begin{equation}
    \begin{gathered}
     \Delta E^{b}(t) \le \overline{\Delta E^{b}}(t) = \lambda E^b(t)
    \end{gathered}
        \label{eq:max_energy_dev}
\end{equation}

For energy calculations of frequency containment reserves in \gls{ems} algorithms, one can assume that $\mathcal{X} \le r_{ss}, \; \forall \; t \; \in \; [t_0,t_0 + T]$ when the system is frequency stable~\cite{alves_sufficient_2021}, this simplification resulting in \cref{eq:energy_bound}.
Remark that, on one hand, $r_{ss} \le \mathcal{X} \le r_{tr}$ during the arrest and rebound periods of frequency containment.
On the other hand, those periods combined last only a couple of seconds, while $T$ may vary from 5 minutes to 15 minutes in typical \glspl{ems}.
The integration error would therefore be minimal when adopting this simplification.
The latter however allows $\widehat{\Delta E^{b}} \le \overline{\Delta E^{b}}(t)$ to be enforced in the optimization problem, as described later in \cref{subsubsec:chance_optProb}, and to satisfy \cref{eq:max_energy_dev}, which avoids excessive deviations of the \gls{soe} from its optimal dispatch.

\begin{equation}
    \begin{gathered}
      \Delta E^{b}(t) = \Delta E^{b}(\mathcal{X}) =  \int\displaylimits_{t_0}^{t_0+T} \left( M_b \dot{\mathcal{X}} + D_b \mathcal{X} \right) \mathrm{d}t = \\
    M_b \mathcal{X} \Bigr|^{t_0+T}_{t_0} + D_b \int\displaylimits_{t_0}^{t_0+T} \mathcal{X} \mathrm{d}t \le M_b r_{tr} + D_b r_{ss} T = \widehat{\Delta E^{b}}
    \end{gathered}
        \label{eq:energy_bound}
\end{equation}


\subsubsection{Adaptive Uncertainty Quantification}
\label{subsubsec:qrf}

Load uncertainty is, in the proposed method, evaluated not only in rapidly varying loads but also in the intermittency of non-dispatchable power sources, being the net load variation defined as the combination of two random variables $P^{p}$, where $p=\left \{\ell,w\right \}$ corresponds to load and non-dispatchable sources respectively).
The time-varying \gls{cdf} for each random variable and predication lead time presented in \cref{eq:F_hat} can be inferred when using ensembles of quantile values following the method in \cite{chapaloglou_data-driven_2022}, where $\bm{x}_r(t)$ is the regressor composed only by lagged values of the corresponding variable.
The net load as a function of time is represented as $\xi_{t+k}(t) = P^{\ell}_{t+k}(t)-P^{w}_{t+k}(t)$, where the two random variables are combined, resulting in the adaptive probabilistic forecasting capability which will be exploited in the construction of the security constrained energy management algorithm described in \cref{subsubsec:chance_optProb}.

\begin{equation}
    \begin{gathered}
        \hat{F^p}^{-1}_{t+k|t} \left( P^{p}_{t+k} \mid \bm{x}_r(t) \right) =  \left \{ Q^{p}_{\tau,k}\left ( \bm{x}_r(t) \right) \right \}_{\tau \in [0,1]}  \Rightarrow \\
        P^{p}_{t+k} \sim \hat{F^p}_{t+k|t} \left( P^{p}_{t+k} \mid \bm{x}_r \right) \; \forall \; k \; \in \; \mathcal{K}.
    \end{gathered}
    \label{eq:F_hat}
\end{equation}


\subsection{Probabilistically Constrained EMS}
\label{subsubsec:chance_optProb}

This section presents how the proposed \gls{ems} algorithm integrates into a single optimization problem including the various objectives and multiple time-scale requirements described in \cref{subsec:pfc,subsec:optProb}.
The system dynamics are first expressed in a hybrid state space system to achieve this integration as shown in \cref{eq:x_lus}, where $\mathbf{x}_{+}$ is the state at the next discrete time, $\mathbf{u}$ the corresponding control inputs and $\bm{A},\bm{B}$ the corresponding system matrices.
Note that $x^{SoC} = E^b / \bar{E}^b \in [\underline{x}^{SoC}, \bar{x}^{SoC}]$ and $\mathbf{x}_{1:N_g}^{gt} = \{ x_{g}^{gt} \}_{1:N_g} \in \{ 0,1 \}$.

\begin{equation}
    \begin{gathered}
    \mathbf{x}_{+} = \bm{A}\mathbf{x}+\bm{B}\mathbf{u}
    \end{gathered}
    \label{eq:x_lus}
\end{equation}

The optimization problem presented in \cref{eq:optProb_general} is then formulated and solved in each discrete time step $t$, where $\bm{\delta} = \{ \delta^i_k \}_{1:K}$ is a random multi-sample, $\bm{\xi} = \{ \xi_{k} \}_{0:K-1}$ is the deterministic mean net load forecast including the measured value $\xi_0$ at $t=t_0$, and $\bm{P}^{b}_{nl} = \{  P^{b}_{nl} \left ( \delta_{k+1} | \xi_{k} \right )  \}_{0:K-1}$ is the net load perturbation for the whole prediction horizon as a function of the sample and the mean forecast. 

\begin{equation}
    \begin{aligned}
        \mathcal{P:} \;\; & \min_{\mathbf{u},\mathbf{v},\mathbf{z}}\{ \mathcal{F}(\mathbf{\mathbf{x}_{+},\mathbf{z},\mathbf{u},\mathbf{v}};\bm{\xi}) \} \\
        \text{where,} \;\;
         & \mathcal{F} \coloneqq \underbrace{\mathcal{J}(\mathbf{x}_{+})}_{\text{states cost}} + \underbrace{\mathcal{J}(\mathbf{z})}_{\text{operation cost}} + \underbrace{\mathcal{J}(\mathbf{u})}_{\text{control cost}} + \underbrace{\mathbf{w}^T \mathbf{v}}_{\text{reserves cost}} \\
        \text{s.t.} \;\; & \mathcal{C}(\mathbf{\mathbf{x}_{+},\mathbf{z},\mathbf{u},\mathbf{v}},\xi) \preceq \bm{0},\\
        &  \mathbb{P} ( \bm{\delta} \in \Delta  \; | \; \abs{\bm{\xi} - \bm{\delta}} = \bm{P}^{b}_{nl} \preceq \\
        &  \min \{ \mathcal{D}r_{ss}(1-r_{tr}), \mathcal{M}\bar{\gamma} \} \mathbb{1}_{ \left \{ 1:K \right \}} ) \ge 1 - \epsilon.
    \end{aligned}
    \label{eq:optProb_general}
\end{equation}

In the cost function $\mathcal{F}$, $\mathcal{J}(\mathbf{x}_{+})$ represents the cost of having generators online, $\mathcal{J}(\mathbf{z})$ penalizes the deviation from optimal operating conditions of online generators and the \gls{ess}, $\mathcal{J}(\mathbf{u})$ captures the start up cost of generators, and finally $\mathbf{w}^T \mathbf{v}$ accounts for costs of activating reserves for frequency control.
The decision variables $\mathbf{v} =[\bm{D}_g, D_b, M_b]^T$ used for frequency reserves are weighted by $\mathbf{w}$ so that different contributions can be assigned to online generators and the \gls{ess} to ensure the system frequency stability.

The term $\mathcal{C}(\cdot)$ encapsulates all the constraints related to the operation of the energy system, as described in \cite{chapaloglou_data-driven_2022}. 
The last expression in \cref{eq:optProb_general} is a chance constraint that, given the estimated distributions from \cref{eq:F_hat}, relates potential instantaneous perturbations for a net load value $\xi_{k}$ and the next step sampled net load $\delta^i_{k+1}$.
\Cref{eq:Dmin,eq:Mmin} are, in other words,  risk-constrained by $\epsilon$, which act as a mechanism to leverage cases with rather pessimistic prediction intervals resulting from poor uncertainty range estimation.

The chance constraint complicates, however, the optimization problem. 
Note that $\mathcal{P}$ is a stochastic \gls{minlp} and therefore non-convex, so a standard \textit{scenario approach} cannot be applied.
Following \cite{vrakopoulou_probabilistic_2013,hreinsson_stochastic_2015}, a probabilistic set composed of a finite number of samples $\bm{\delta^i} \; \in \Delta^N$ is computed as in \cref{eq:N_scenarios}, being $\beta$ and $e$ user defined parameters that tighten the non-violation of the original chance constraint.
Note that the number of random variables is $2|\mathcal{K}|$ because two sources of uncertainty exist (load and renewable power injection) for the whole prediction horizon.
The chance constraint can however be replaced by the deterministic set of linear inequalities presented in \cref{eq:chance_cstr_reform}, which
transforms the stochastic problem $\mathcal{P}$ into a robust one, where the set $\Delta^N$ encapsulates the same risk-volume $\epsilon$ as the original chance constraint.
When solving the corresponding robust program, solutions are not only feasible for the initial $\mathcal{P}$ but also satisfy the same probabilistic guarantees \cite{margellos_road_2014}.

\begin{align}
    N &\ge \frac{1}{\epsilon}\frac{e}{e-1} \left ( \ln{\frac{1}{\beta}} + 4|\mathcal{K}| -1 \right  ) \label{eq:N_scenarios} \\
    \abs{\bm{\xi} - \bm{\delta^i}} = \bm{P}^{b}_{nl} &\preceq \min \{ \mathcal{D}r_{ss}(1-r_{tr}), \mathcal{M}\bar{\gamma} \} \mathbb{1}_{ \left \{ 1:K \right \}} \label{eq:chance_cstr_reform} \\
    & \forall \; \bm{\delta^i} \; \in \Delta^N \nonumber
\end{align}

\subsection{Proposed Convex Reformulation}
\label{subsec:milp_reform}

It is not feasible, on one hand, to implement the constraints expressed in \cref{eq:Dmin,eq:Dsum,eq:Pglim} directly into a \gls{milp} formulation because of the bi-linear terms $x_{g}^{gt} D_{g}$.
It is possible, on the other hand, to take advantage of the following observation: with $N_{g}$ generators, there are exactly $2^{N_g}$ different possible system configurations depending on the values of $\mathbf{x}_{1:N_g}^{gt} \in \{ 0,1 \}$.
The status variable $x_{g,t}^{gt}$ takes a particular value (0 or 1) for each generator and configuration, being therefore feasible to represent it by binary strings of length $N_g$.
The feasible system configurations are then enumerated by taking all the possible permutations, which can be gathered in a data table $\bm{A}_{cf} \; [2^{N_g } \times N_g ]$.
The variables $x_{g,k}^{gt}, \; \forall \; k \in \mathcal{K}$ can, in this way, be treated as constants given a selected system configuration, being indicator variables $b_j$ introduced
to identify the configuration selected and associate it with the status of the generators by the set of constraints in \cref{eq:a_b1,eq:a_b2}.
\Cref{eq:Dsum} can, as consequence, be linearized using $\bm{A}_{cf}$ as in \cref{eq:D_lin}.

\begin{align}
    x_{g,k}^{gt} &\ge b_j, \; \{(j,g) \in \mathcal{J} \times \mathcal{N}_g \mid \left[A_{cf}\right ]_{j,g} = 1 \} \label{eq:a_b1} \\
    x_{g,k}^{gt} &\le 1- b_j, \; \{(j,g) \in \mathcal{J} \times \mathcal{N}_g \mid \left[A_{cf}\right ]_{j,g} = 0 \} \label{eq:a_b2} \\
    \bm{\mathcal{D}} &= \bm{A}_{cf} \bm{D}_g + D_b \mathbb{1}_{ \left \{ 1:J \right \}}
    \label{eq:D_lin}
\end{align}

\Cref{eq:Dmin,eq:Mmin} can also be reformulated using the sampled net \gls{apd} terms from \cref{eq:chance_cstr_reform}.
This is presented in \cref{eq:Dmin_Mmin_ineq}, where $M_B$ is a big-M value, $\bm{b}_k, \bm{\mathcal{D}}_k$ correspond to each lead time of the prediction horizon ($k \in \mathcal{K}$) assuming that $M_g$ are constant values.
The different possible configurations are restricted by a type 1 \gls{sos} constraint as in \cref{eq:b_j_sos1}.

\begin{align}
    \bm{P}^{b}_{nl} \left ( \delta^i_{k+1} | \xi_k \right ) + M_B \bm{b}_k &\preceq \bm{\mathcal{D}}_k r_{ss} \left (  1-r_{tr} \right) + M_B  \mathbb{1}_{ \left \{ 1:J \right \}} \nonumber \\
    \frac{1}{\bar{\gamma}}\bm{P}^{b}_{nl} \left ( \delta^i_{k+1} | \xi_k \right ) &\le M_{g}  \mathbb{1}^T_{ \left \{ 1:N_g \right \}} \bm{x}_{k}^{gt} + M_{b,k} \label{eq:Dmin_Mmin_ineq} \\
    & \forall \; \delta^i_k \; \in \Delta^N, \;\; \forall \; k \in \mathcal{K} \nonumber \\
    \mathbb{1}_{ \left \{ 1 : J \right \}}^T \cdot \bm{b}_k &= 1,
      \;\; \forall \; k \in \mathcal{K} \label{eq:b_j_sos1}    
\end{align}

\Cref{eq:Pglim} is similarly linearized using $\bm{A}_{cf}$ and $M_B$ as in \cref{eq:Pgt_ineq_reform_A,eq:Pgt_ineq_reform_B} where $S_b$ is the system's base power.

\begin{equation}
    \begin{gathered}
        \bm{P}_{k}^{gt} + M_B \bm{b}_k \preceq \left ( \bar{P}^{gt} - D_{g,k} \cdot r_{tr} \cdot S_b \right )  + M_B  \mathbb{1}_{ \left \{ 1:J \right \}},  \\
        \left ( \underline{P}^{gt} + D_{g,k} \cdot r_{tr} \cdot S_b \right )  +  M_B \bm{b}_k \preceq  \bm{P}_{k}^{gt} + M_B\mathbb{1}_{ \left \{ 1:J \right \}} \\
         \;\; \forall \; k \in \mathcal{K}, \; \{(j,g) \in \mathcal{J} \times \mathcal{N}_g \mid \left[A_{cf}\right ]_{j,g} = 1 \}
    \end{gathered}
    \label{eq:Pgt_ineq_reform_A}
\end{equation}
\begin{equation}
    \begin{gathered}
        \bm{P}_{k}^{gt} + M_B \bm{b}_k \preceq M_B\mathbb{1}_{ \left \{ 1:J \right \}}, \\
        M_B \bm{b}_k \preceq \bm{P}_{k}^{gt} + M_B\mathbb{1}_{ \left \{ 1:J \right \}}, \\
        \; \forall \; k \in \mathcal{K}, \; \{(j,g) \in \mathcal{J} \times \mathcal{N}_g \mid \left[A_{cf}\right ]_{j,g} = 0 \}
    \end{gathered}
    \label{eq:Pgt_ineq_reform_B}
\end{equation}

The \gls{ess} reserves from \cref{eq:Pblim} are reformulated in terms of discharging and charging power as in \cref{eq:Pch_Pdis_ineq_reform}, where $s_k$ is the discharging binary indicator variable.

\begin{equation}
    \begin{gathered}
        P_{k}^{dis} \le s_k\bar{P}^{b},  \\
        P_{k}^{dis} \le \bar{P}^{b} - \left ( D_{b,k} \cdot r_{tr} + M_{k,b} \cdot \bar{\gamma} \right ) \cdot S_b, \\
        P_{k}^{ch} \le (1-s_k)\bar{P}^{b},  \\
        P_{k}^{ch} \le \bar{P}^{b} - \left ( D_{b,k} \cdot r_{tr} + M_{k,b} \cdot \bar{\gamma} \right )\cdot S_b, \; \forall \; k \in \mathcal{K}
    \end{gathered}
    \label{eq:Pch_Pdis_ineq_reform}
\end{equation}

The bound derived in \cref{eq:energy_bound} can be easily integrated to $\mathcal{P}$ through the linear constraints in \cref{eq:ess_bound_reform}, where $\nu =  \frac{r_{ss}S_b}{3600 \bar{E}^b}$.

\begin{equation}
    \begin{gathered}
       \nu \left ( M_{b,k} + D_{b,k}T \right ) \le  \\
        \min \{ \bar{x}^{SoC} - x^{SoC}_{k-1}, x^{SoC}_{k-1} - \underline{x}^{SoC}, \lambda x^{SoC}_{k} \}, \;
         \;\; \forall \; k \in \mathcal{K}
    \end{gathered}
    \label{eq:ess_bound_reform}
\end{equation}

\Cref{eq:optProb_general} may therefore be expressed as a deterministic \gls{milp} robust program and solved efficiently by the \gls{ems} at each time iteration, as $\mathcal{F}(\cdot)$ and $\mathcal{C}(\cdot)$ can be expressed as linear combinations of the optimization variables when including the reformulations proposed in \cref{eq:chance_cstr_reform,eq:a_b1,eq:a_b2,eq:D_lin,eq:Dmin_Mmin_ineq,eq:b_j_sos1,eq:Pgt_ineq_reform_A,eq:Pgt_ineq_reform_B,eq:Pch_Pdis_ineq_reform,eq:ess_bound_reform}.

\section{Simulations}
\label{sec:simulations}

The various components of the proposed methodology are demonstrated in this section, namely the adaptive uncertainty quantification and the integration of frequency stability constraints with bounded use of the \gls{ess} stored energy in the optimal scheduling.
To achieve this goal, the case study of a wind-powered offshore \gls{og} platform is presented, where a reference operation time period is considered involving both regions of smooth, low magnitude and sudden, large net load variations.
The effectiveness of the proposed methodology is then demonstrated with time domain simulations on the scheduling time scale.
The optimization problem is solved with \textit{Gurobi 9.1.0} in a 28 physical core multi-node cluster with Intel(R) Xeon(R) CPU E5-2690 v4 @ 2.60 Hz and 25 GB RAM. 
The solution time of \cref{eq:optProb_general} using the proposed formulation is well below 15 minutes, which is assumed here as the minimum threshold for real-time power system scheduling.

\subsection{Capabilities of adaptive uncertainty quantification}
\label{subsec:qrf_load_step}

\begin{figure}[t!]
    \centering

    \subfloat[$t+k|t_0-3$]{\includegraphics[width=0.7\columnwidth]{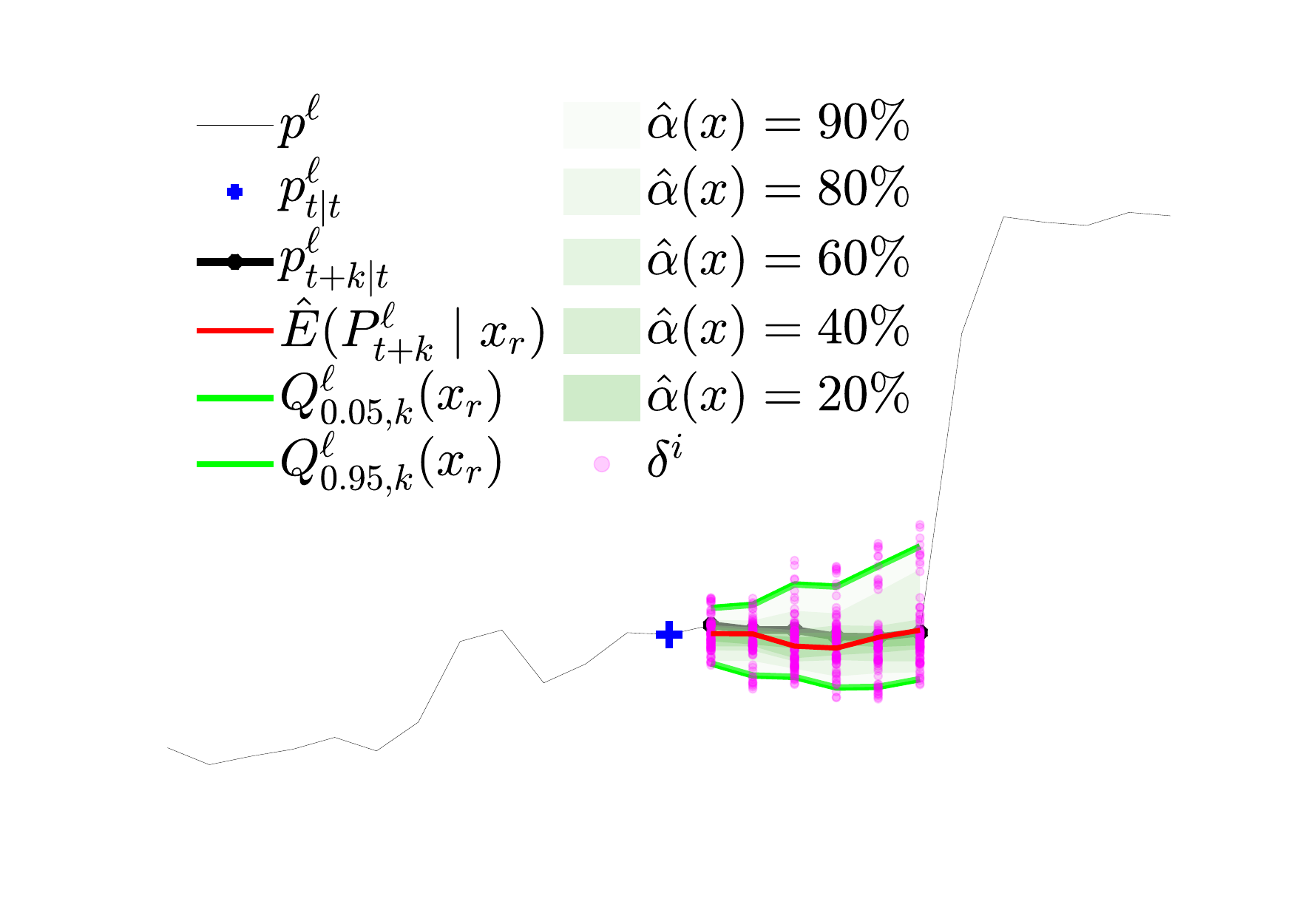}%
    \label{fig:load_k_0}}
    \\
    \hfill
    \subfloat[$t_0-2$]{\includegraphics[width=0.5\columnwidth]{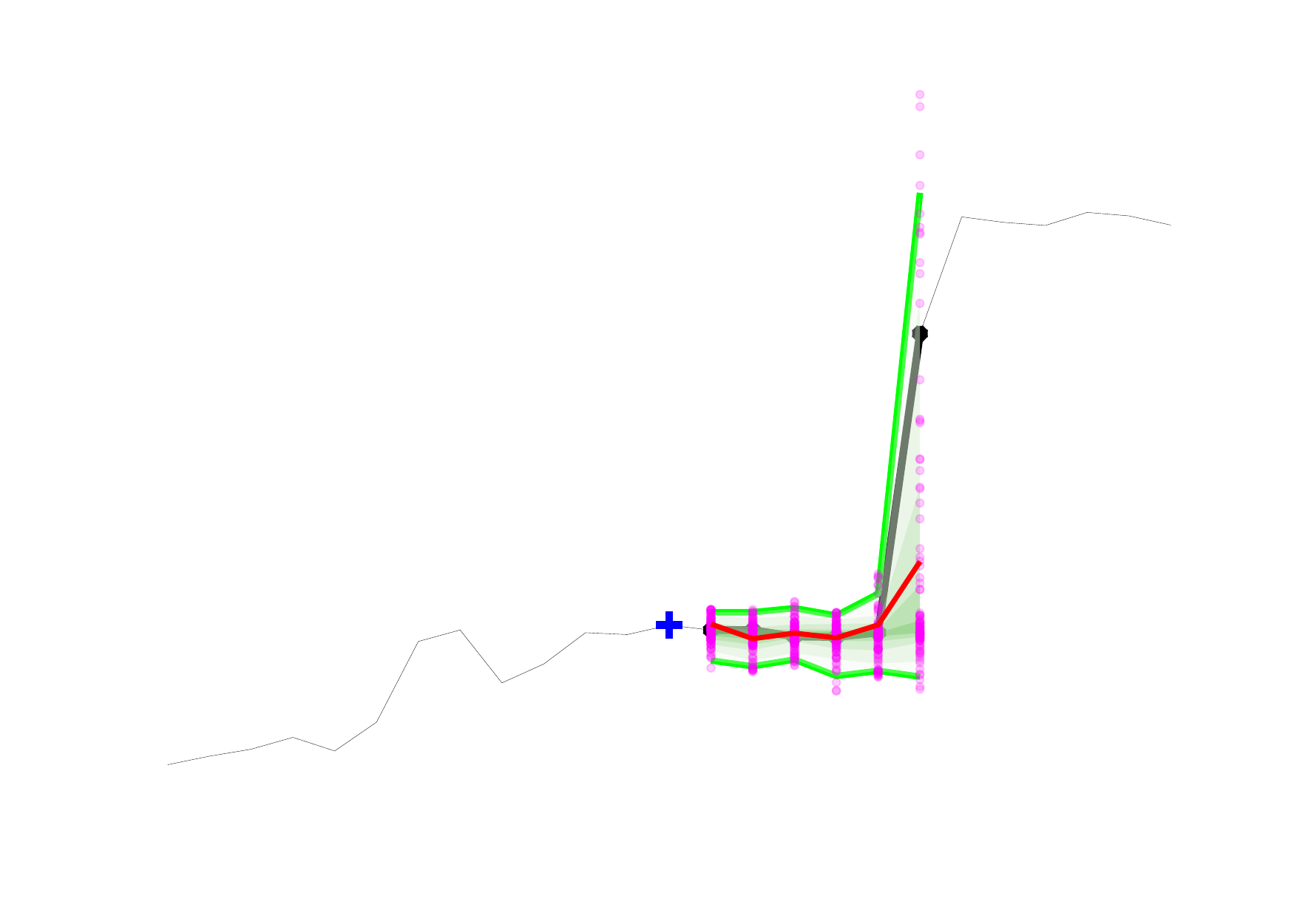}%
    \label{fig:load_k_2}}
    \hfill
    \subfloat[$t_0 -1$]{\includegraphics[width=0.5\columnwidth]{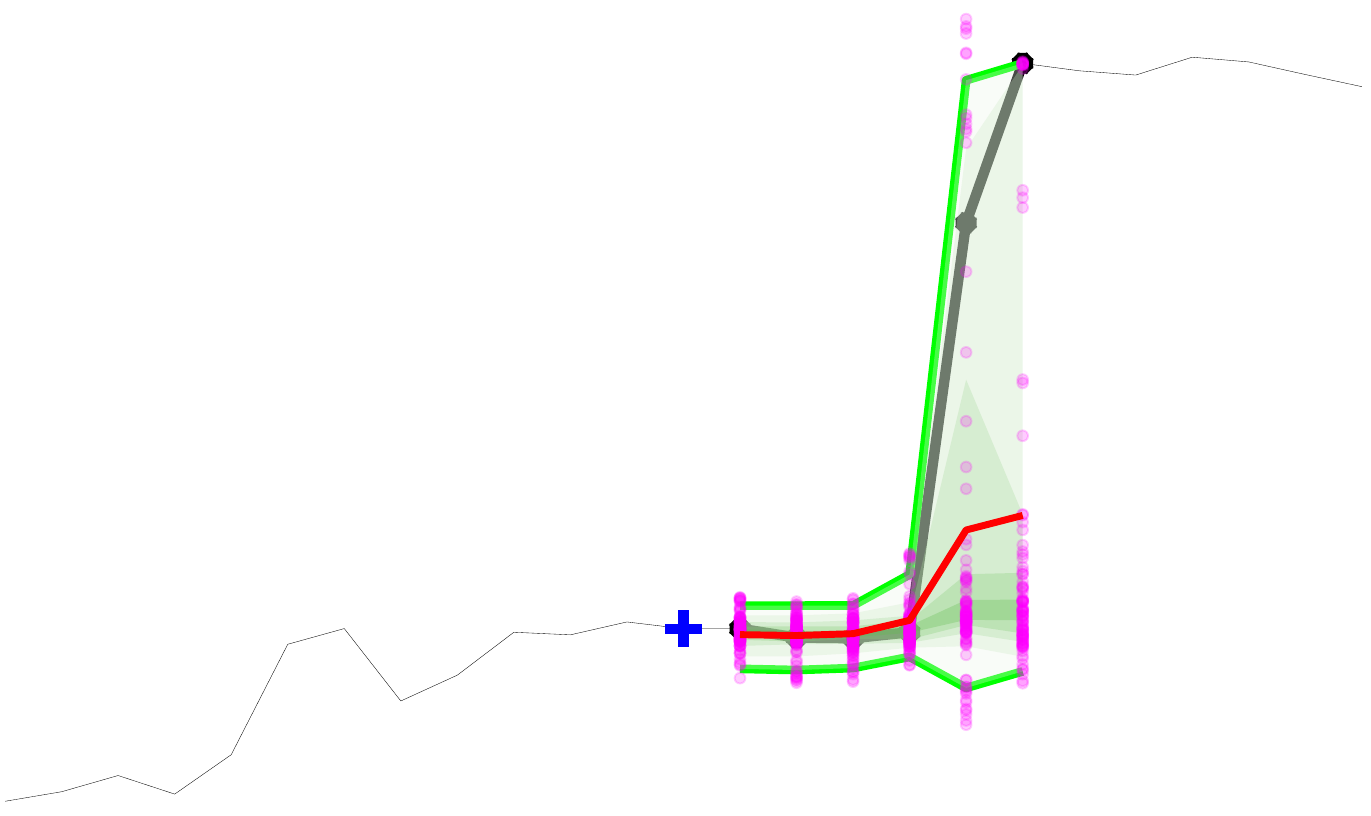}%
    \label{fig:load_k_3}}
    \\
    \subfloat[$t_0$]{\includegraphics[width=0.7\columnwidth]{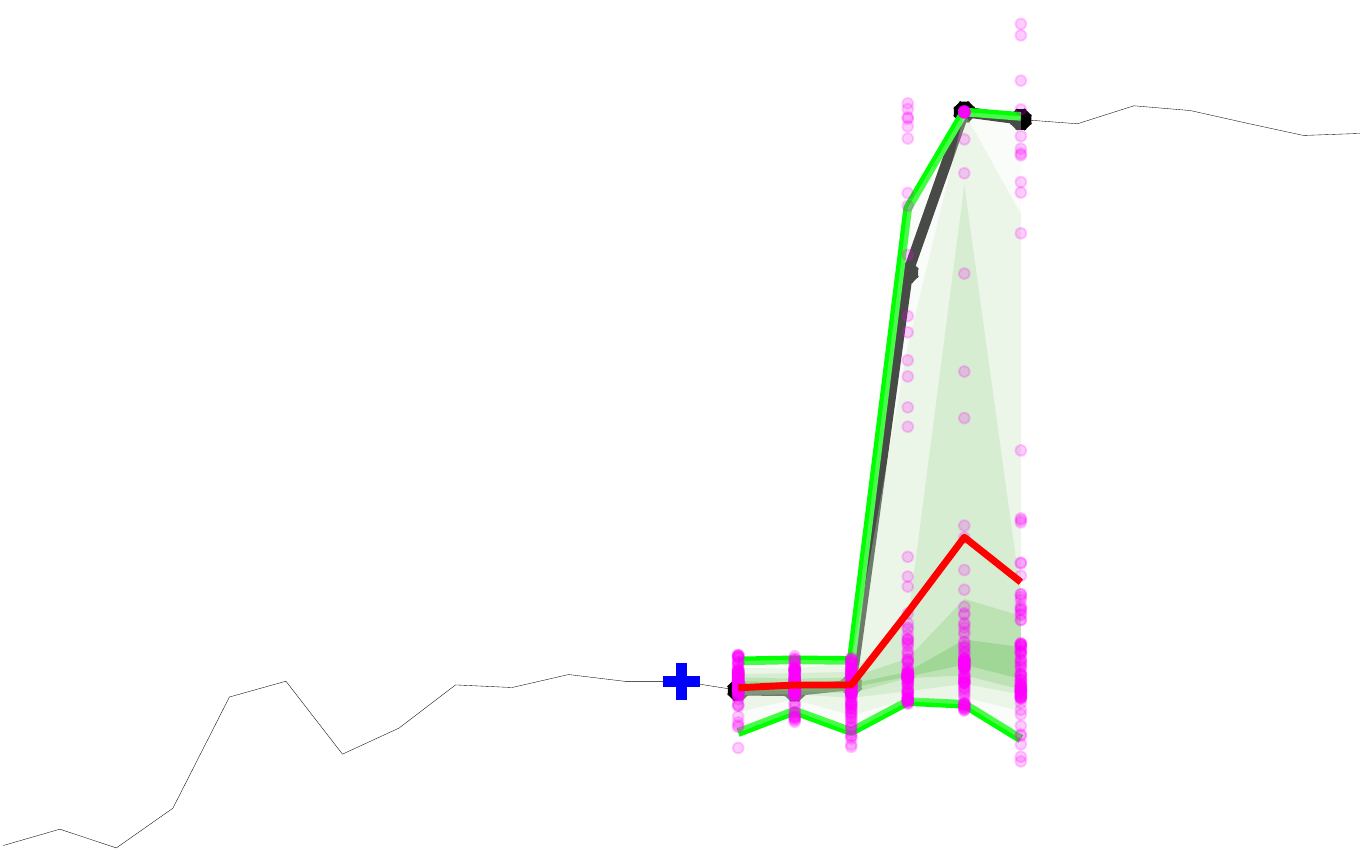}%
    \label{fig:load_k_1}}
    \\
    \subfloat[$t_0 + 1$]{\includegraphics[width=0.33\columnwidth]{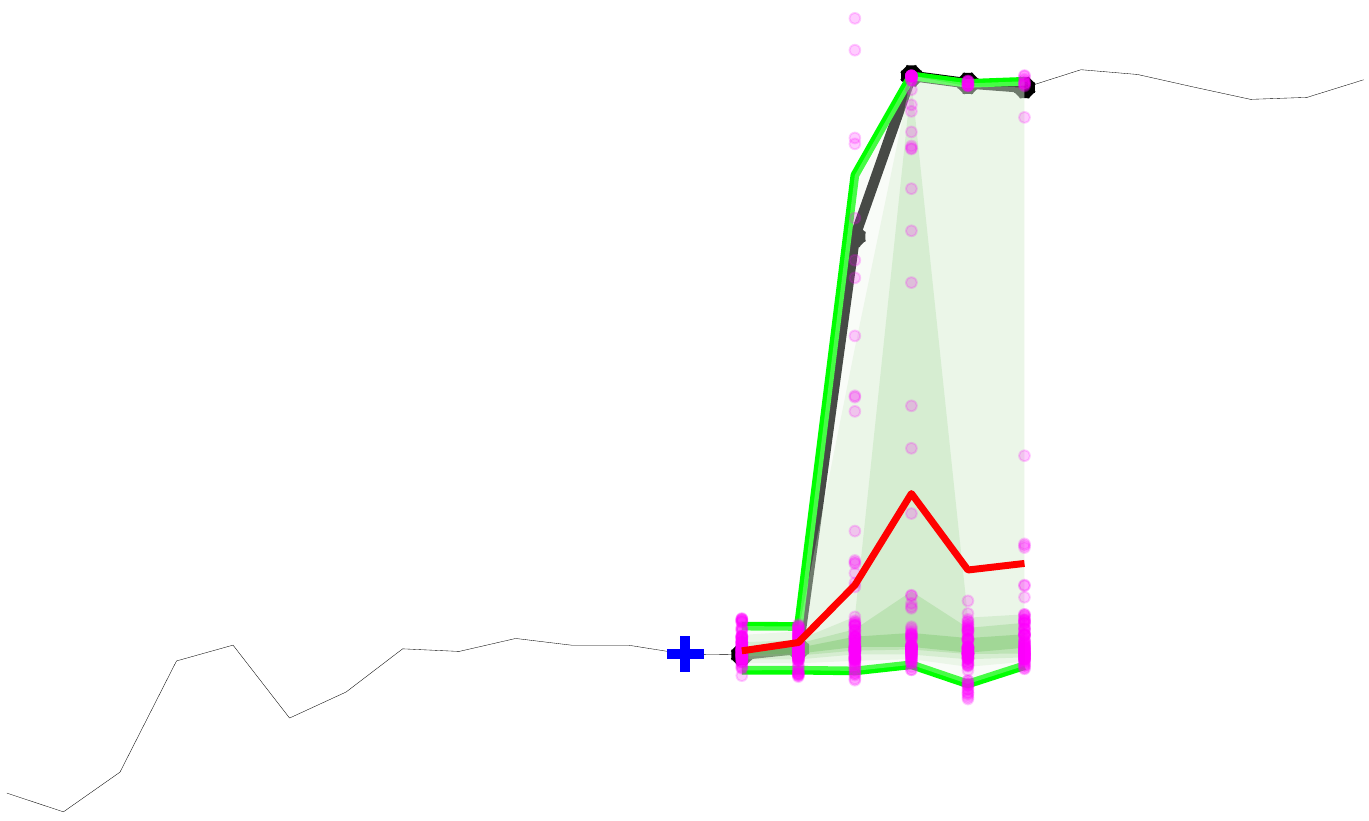}%
    \label{fig:load_k_4}}
    \hfill
    \subfloat[$t_0 + 2$]{\includegraphics[width=0.33\columnwidth]{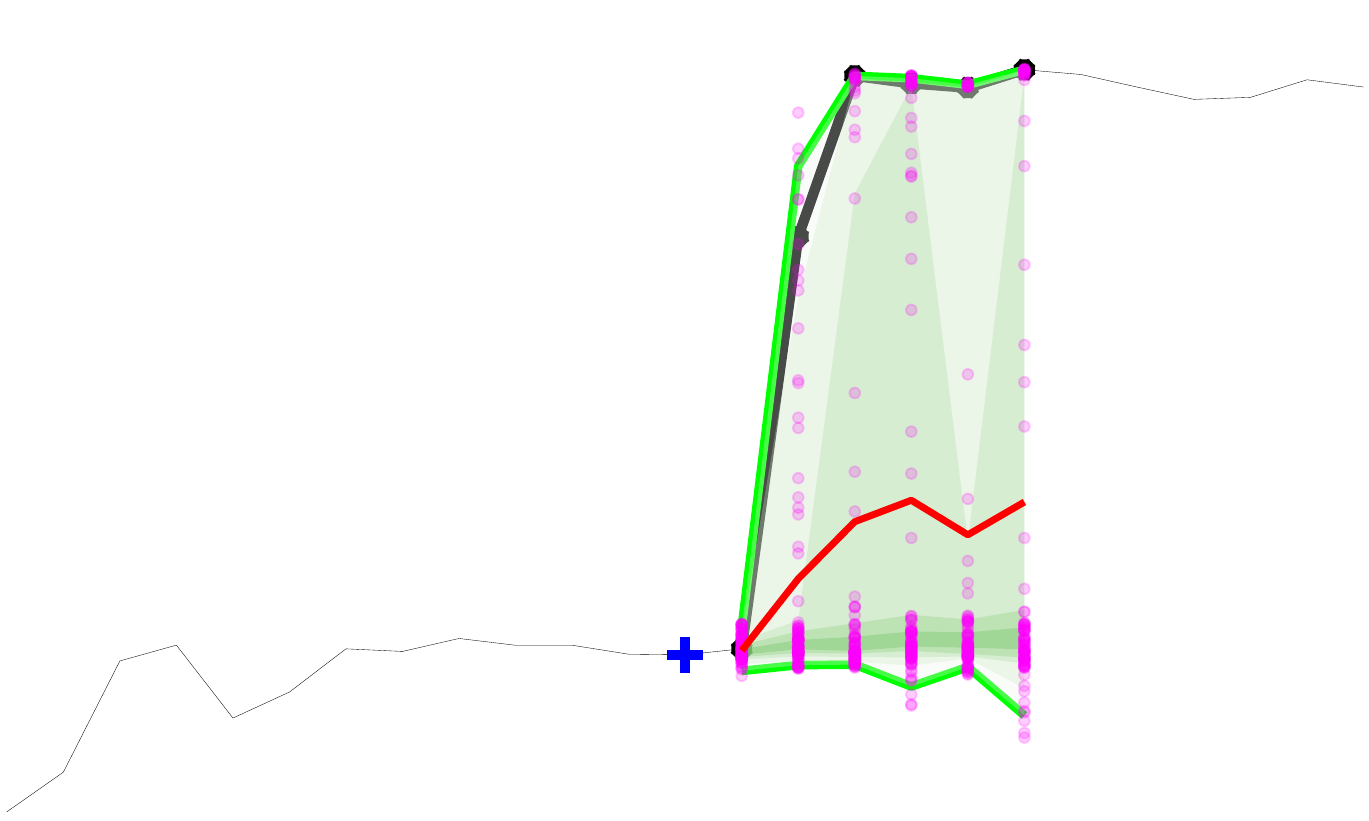}%
    \label{fig:load_k_5}}
    \hfill
    \subfloat[$t_0 + 3$]{\includegraphics[width=0.33\columnwidth]{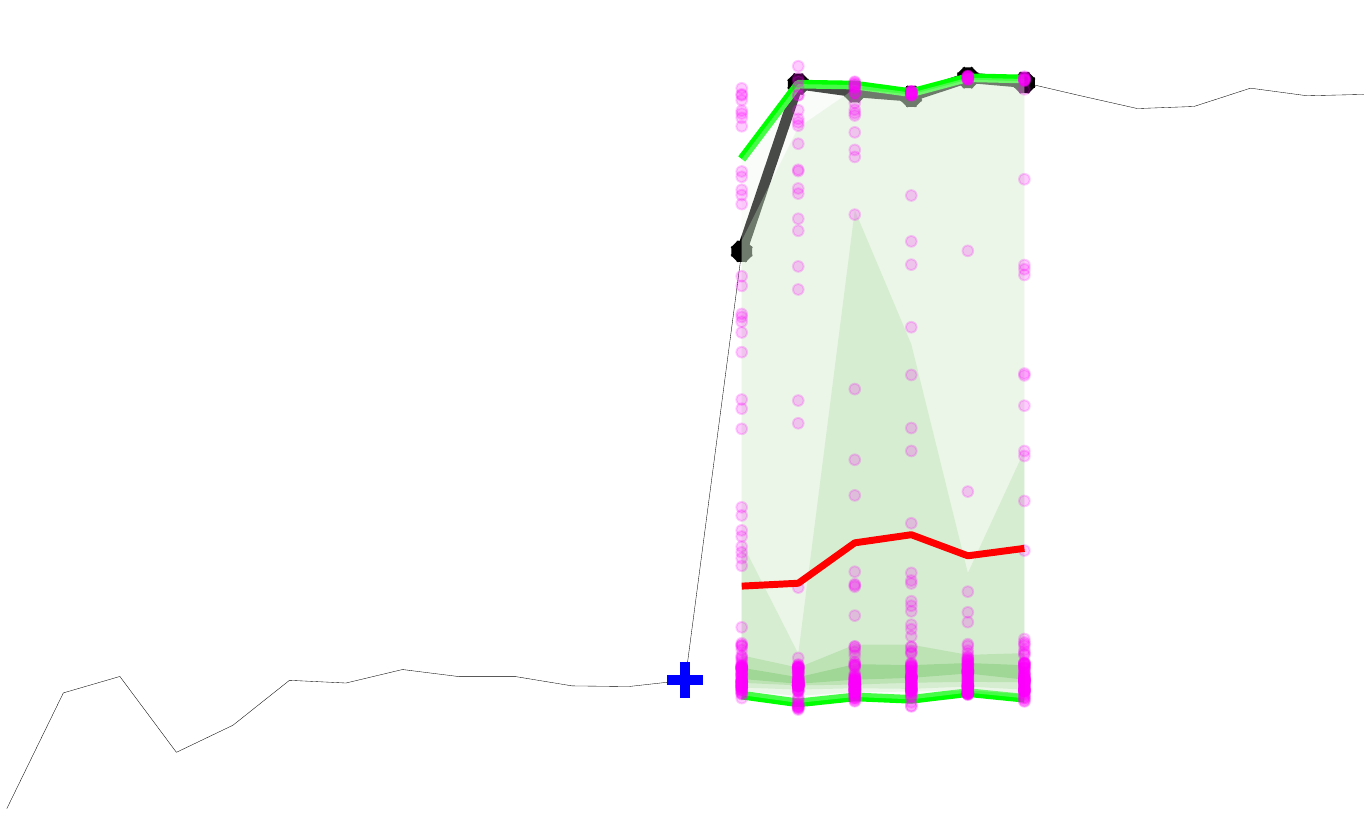}%
    \label{fig:load_k_6}}
    \caption{Demonstration of adaptive uncertainty quantification by using auto-regressive probabilistic forecasting for the load time-series. A case for a sudden step-like variation is presented for consecutive lead times (\cref{fig:load_k_0,fig:load_k_1,fig:load_k_2,fig:load_k_3,fig:load_k_4,fig:load_k_5,fig:load_k_6}). By updating the estimated prediction intervals it is possible to capture the sudden load variation and draw samples (purple dots) that span an appropriate range of values.}
    \label{fig:load_step}
\end{figure}

A representative example of how the adaptive uncertainty quantification framework works is illustrated in \cref{fig:load_step}, where a case of a sudden \gls{apd} was selected since these are the most interesting from a power imbalance perspective.
The actual load values are presented with the solid black line and the blue cross indicates the time instant at which a probabilistic forecast is issued.
From that moment and for a given prediction horizon ($k \in \mathcal{K}$), the expected values of the prediction are plotted in red dashed line along with the prediction intervals around them in green at various quantile levels ($\hat{\alpha}$).
Samples generated by the estimated \glspl{cdf} are also plotted as purple dots.

The proposed adaptive uncertainty quantification algorithm generates samples that better describe the size of an \gls{apd}.
Observe that, as the blue cross moves forward in time, the prediction intervals adapt to capture the irregular event of the sudden load increase.
Note also that, at the initial time (\cref{fig:load_k_0}), the prediction intervals are narrow and all the sampled values fall close and around the actual load values.
From the next moment and later (\cref{fig:load_k_1,fig:load_k_2,fig:load_k_3,fig:load_k_4,fig:load_k_5,fig:load_k_6}), however, the uncertainty increases and the prediction intervals expand to capture the possibility of an irregular sudden load increase.
The deterministic forecast, in contrast, fails to capture the variation adequately, because it is dominated by the inertia of past values, a common drawback of auto-regressive models.

\subsection{Effect of optimal and bounded frequency support from ESS}
\label{subsec:load_step}
To demonstrate the effect of optimally controlling the \gls{ess} to provide frequency support for an isolated \gls{aps}, a step \gls{apd} is considered at time $t_0'= 2 s$ representing a 0.4 pu load increase from a sudden motor startup, where a single \gls{gt} is on in the platform of the case study.
Simulations were run in Matlab/Simulink 2022a using the model given in \cref{eq:nlSwing}, which do not include delay of actuators.
This simplification facilitate the interpretation of results, not implying however any loss of generality.

\begin{figure}[btp!]
    \centering
    \centerline{\includegraphics[width=0.9\columnwidth]{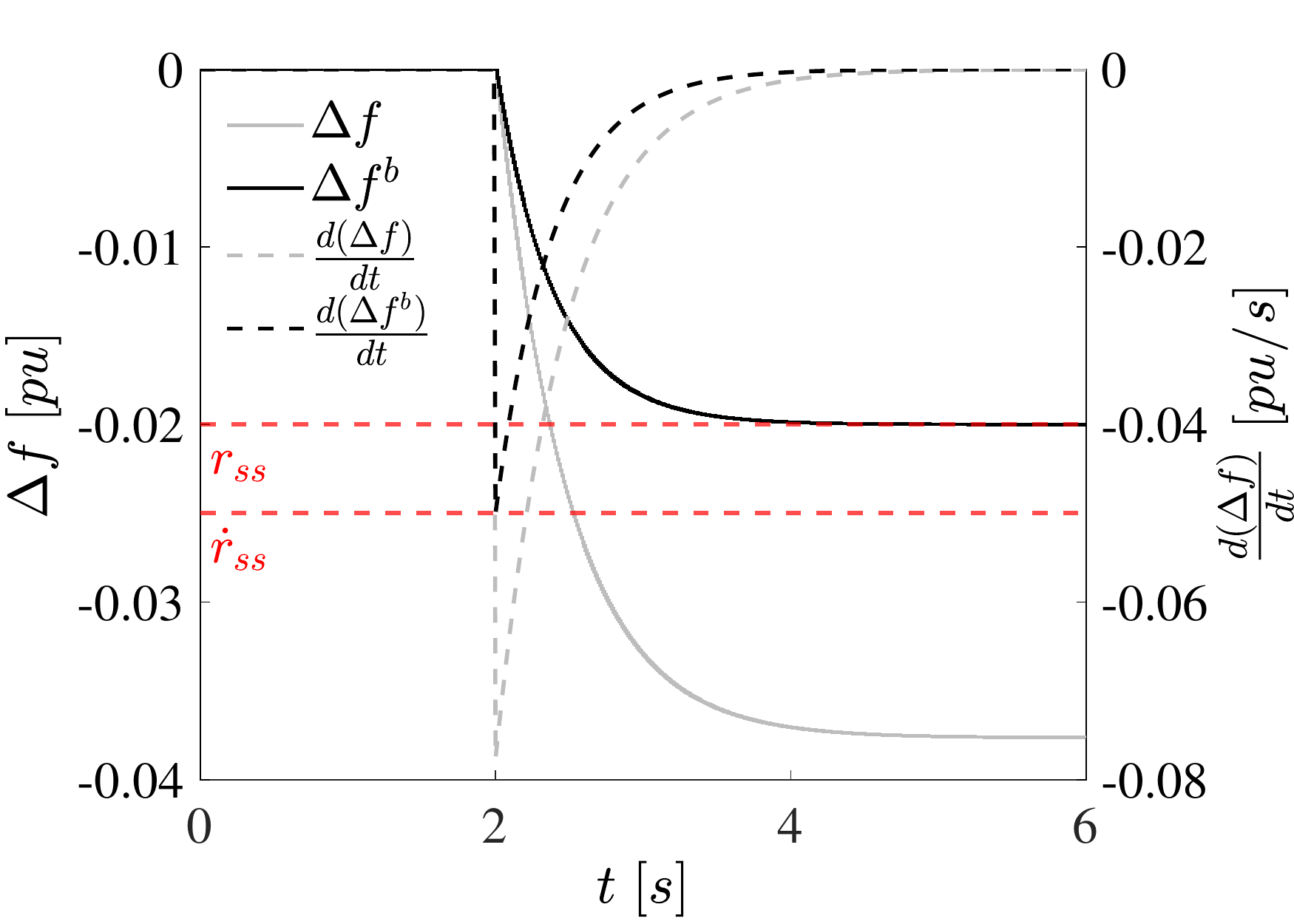}}
    \caption{Effect of the optimally calculated participation of the \gls{ess} in providing frequency support for a step load change when only 1 \gls{gt} is on. In contrast with the case of non participation of the \gls{ess} to frequency regulation (solid and dashed grey lines), the optimally designed virtual inertia and damping results in a frequency response (solid black) and \gls{rocof} (dashed black) that are bounded by their defined limit values.}
     \label{fig:stepPlot}
\end{figure}

Results are illustrated in \cref{fig:stepPlot}, where the frequency deviation (solid lines) and the \gls{rocof} (dashed lines) for two cases are represented.
In the first case (grey lines), the single online \gls{gt} is the only source of primary frequency control, while, in the second case (black lines), the \gls{ess} supports in this task the \gls{gt}, which has the same droop setting as in the previous case.
Observe that the system's response in the first case violates not only the steady-state frequency bound $r_{ss}$ but also the maximum allowable \gls{rocof} ($\Bar{\gamma}$), whereas in the second case both limits are respected.
To respect the defined bound in the first case, the \gls{gt} droop setting must be increased, leading to a larger deviation from the optimal \gls{gt} operating point, decreased efficiency, increased fuel consumption and emissions.
The same droop setting for the \gls{gt}, on the other hand, can be maintained and the $r_{ss}$ threshold observed where an optimal participation of the \gls{ess} in the \gls{pfc} has been decided beforehand by the algorithm using an adaptive droop setting.
Note also the compliance with the \gls{rocof} limit where the \gls{ess} provides virtual inertia.
To respect this limit without the \gls{ess} support would require an additional \gls{gt} on, affecting significantly the overall efficiency of the system.
The proposed algorithm, in other words, employs an adaptive droop and virtual inertia scheme, enabling not only the optimal scheduling of the power system by avoiding an additional \gls{gt} on, but also guaranteeing that the services provided by the \gls{ess} to the grid will not cause it to deviate significantly from its optimal schedule, as the use of stored energy is bounded by \cref{eq:ess_bound_reform}.
As a matter of fact, the calculated energy for frequency support provision by the \gls{ess} to the described disturbance for $T$ (15 minutes) is 0.9179 $MWh$ while the bound for $\lambda=3\%$ is 0.9369 $MWh$.

\subsection{Comparative analysis and effect of bounds}
\label{subsec:compare}
To further demonstrate the capabilities of the proposed \gls{ems}, a reference operation period is simulated considering: \textit{I)} the default \gls{ems} that does not integrate any bounds; \textit{II)} the version that includes frequency variation bounds but not the energy bounds on the \gls{ess}; and \textit{III)} the proposed version which includes all of them. 

\subsubsection{System inertia and damping evolution}
\label{subsubsec:M_n_D}
\begin{figure}[tbp!]
    \centering
    \subfloat[Total system inertia.]{\includegraphics[width=\columnwidth]{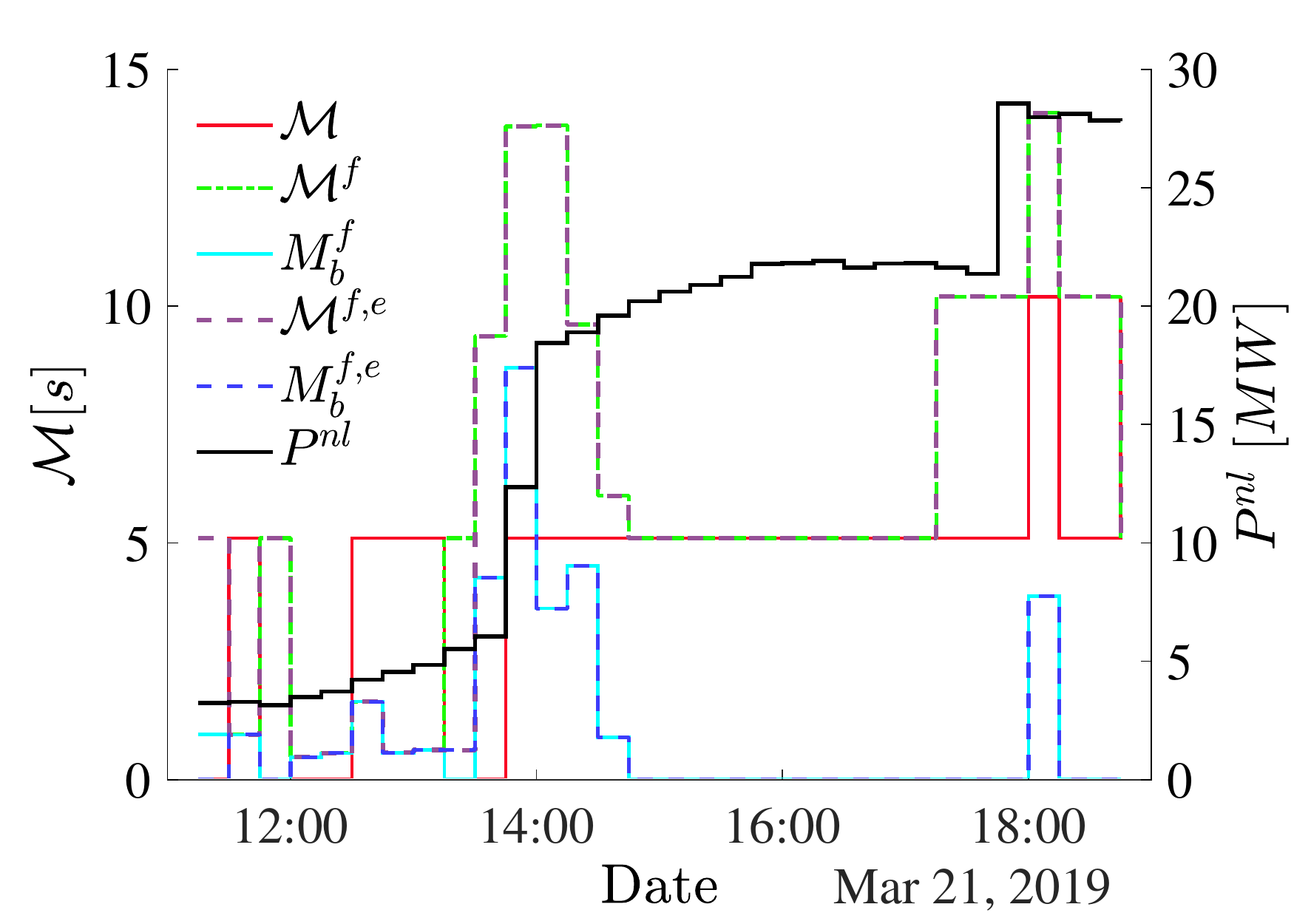}%
    \label{fig:M}}
    \\
    \subfloat[Total system damping.]{\includegraphics[width=\columnwidth]{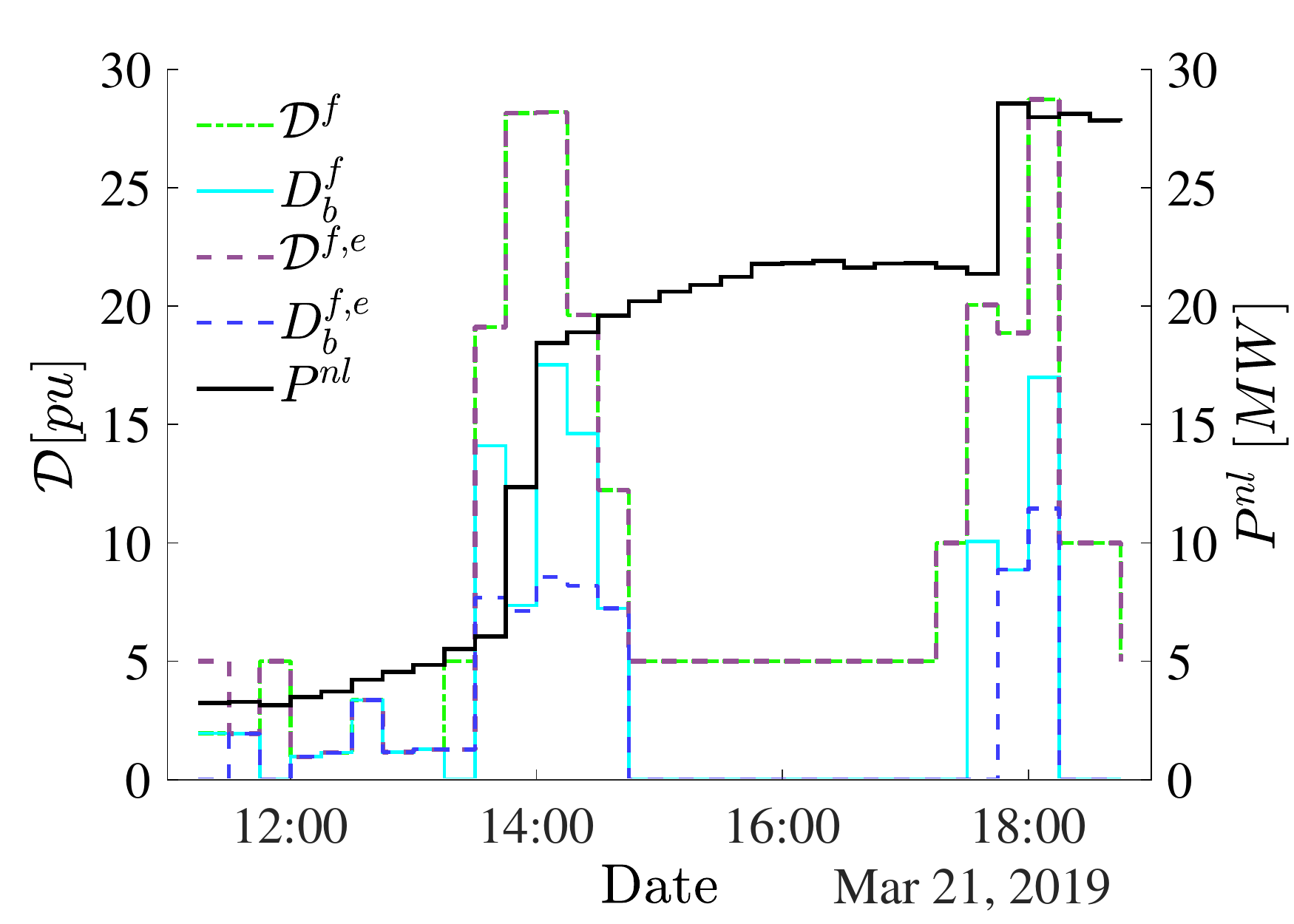}%
    \label{fig:D}}
    \caption{Trajectories of the optimally designed system inertia (top) and damping (bottom) and optimal split between primary control provision by the \glspl{gt} and the \gls{ess} for different bounds considerations. The simulated results are plotted against the net load signal (solid black) for a case where sudden step-like variations occur.}
    \label{fig:M_D}
\end{figure}

The results for all \gls{ems} versions are aggregated in \cref{fig:M_D}, being the system inertia and damping evolution depicted in \cref{fig:M} and \cref{fig:D} along with the net load signal $P_b^{nl}(t)$ (solid black line), during a reference period of 8 hours.
No frequency and \gls{rocof} containment reserves are considered in case \textit{I}, the droop settings not being optimization variables and the \gls{ess} not contributing with virtual inertia.
The system inertia $\mathcal{M}(t)$ (red line in \cref{fig:M}) is therefore just the result of the online \glspl{gt} based on the optimal techno-economic scheduling, not having any virtual component $M_b(t)$. Following the same color notation in both \cref{fig:M} and \cref{fig:D}, the evolution of the system inertia and damping is observed, being the contribution of the \gls{ess} in case \textit{II} denoted by the superscript $f$ (frequency bound) and in case \textit{III} by $f,e$ (frequency and energy bound). 

Two main patterns are evident when observing \cref{fig:M} and \cref{fig:D}.
Firstly, more inertia and damping are assigned for both cases \textit{II} and \textit{III} close to the time instants of the sudden net load variations (around 14:00 and 18:00 correspondingly).
Secondly, the same inertia and damping are noticed during the relatively constant net load conditions (between 14:00 and 18:00).
This effect demonstrates the adaptive capabilities of the proposed \gls{ems} algorithm to assign more or less inertia and damping in correspondence with the anticipated net load variations.
The larger the variations expected, the more secure the system will be by properly deciding its power and energy reserves.
Whether the \gls{ess} is not assigned to frequency control, securing the system for possible net load variations would require additional \glspl{gt} to be on. 
This is confirmed in \cref{fig:M} during both sudden net load variations, where $\mathcal{M}(t) \le \mathcal{M}^f(t)$ (green line) and $\mathcal{M}(t) \le \mathcal{M}^{f,e}(t)$ (magenta line).
Note also that around those instants, $M_b^f(t) \ge 0$ (cyan line) and $M_b^{f,e}(t) \ge 0$ (blue line), meaning that the additional \gls{gt} can be avoided by properly assigning virtual inertia to the \gls{ess}. 
Similar observations are drawn from \cref{fig:D} where $D_b^f(t) \ge 0$ (cyan line) and $D_b^{f,e}(t) \ge 0$ (blue line) around the sudden variation instants and $D_b^f(t) = 0$ and $D_b^{f,e}(t) = 0$ for the rest period.

Remark that the effect of the energy bounds on the \gls{ess} are stronger in the damping terms than in the inertia.
\Cref{fig:M} shows that the signals $M_b^f(t)$ and $M_b^{f,e}(t)$ are almost identical for the whole time period, whereas \cref{fig:D} depicts that $\sup D_b^{f,e}(t) \le \sup D_b^f(t)$.
This means that the peak contribution of the \gls{ess} for case \textit{III} is smaller than the one for case \textit{II}.
It is indeed possible to observe that $D_b^{f,e}(t)$ is almost always less than $D_b^{f}(t)$, further demonstrating that version \textit{III} is more cautious not to overuse the \gls{pfc} of the \gls{ess}.  

The comparison of the different methods (\textit{I}, \textit{II}, \textit{III}) is also quantified through the cumulative results of specified key performance indicators in \cref{table:kpi}.
Including bounds on the \gls{ems} results in slightly higher fuel consumption and operating costs, which is primarily attributed to higher cumulative operational hours of the \glspl{gt}.
This is in agreement with the results presented in \cref{fig:M_D}, as securing the system against possible disturbances may be associated with increased inertia requirements.
It is also noteworthy in \cref{table:kpi} that the incremental cost for including bounds on the use of \gls{ess} stored energy (\textit{III}) compared to case \textit{II} is negligible, meaning that the optimal \gls{gt} scheduling is almost not affected by the inclusion of energy bounds in the \gls{ess}.
The main difference is that the resulting optimal \gls{gt} trajectory for \textit{II} is associated with one less \gls{gt} start up compared to \textit{III} and slightly lower \gls{ess} cycling, reflected on the lower degradation, which is however almost equal in cases \textit{I} and \textit{III}.
\begin{table}[b!]
    \caption{Performance comparison for the whole simulation period.}
    \begin{center}
        \begin{tabular}{ cccc }
    \hline
    \multirow{2}{*}{\textbf{Performance indicator}} & \multicolumn{3}{c}{\textbf{Method}} \\
    & \textit{I} & \textit{II} & \textit{III}  \\
    \hline

 fuel consumption (ON \glspl{gt}) [kg]           & 29,695    & 29,929 & 30,131 \\
    fuel costs (ON \glspl{gt})  [\EUR{}]         & 8,846     & 8,916 & 8,976 \\
    \glspl{gt} ON time              [$N_g \times T$]          & 27        & 31 & 31 \\
    \glspl{gt} startup times            [-]      & 5         & 4 & 5 \\
    \gls{ess} degradation [\%]                   & 0.164     & 0.148 & 0.165 \\
    

    \hline
    \hline
\end{tabular}
    \end{center}
    \label{table:kpi}
\end{table}


\subsubsection{State of charge evolution and energy bounds effect}
\label{subsubsec:SoC_n_Bounds}

\begin{figure}[tbp!]
    \centering
    \subfloat[Without bounds on the energy deviation.]{\includegraphics[width=\columnwidth]{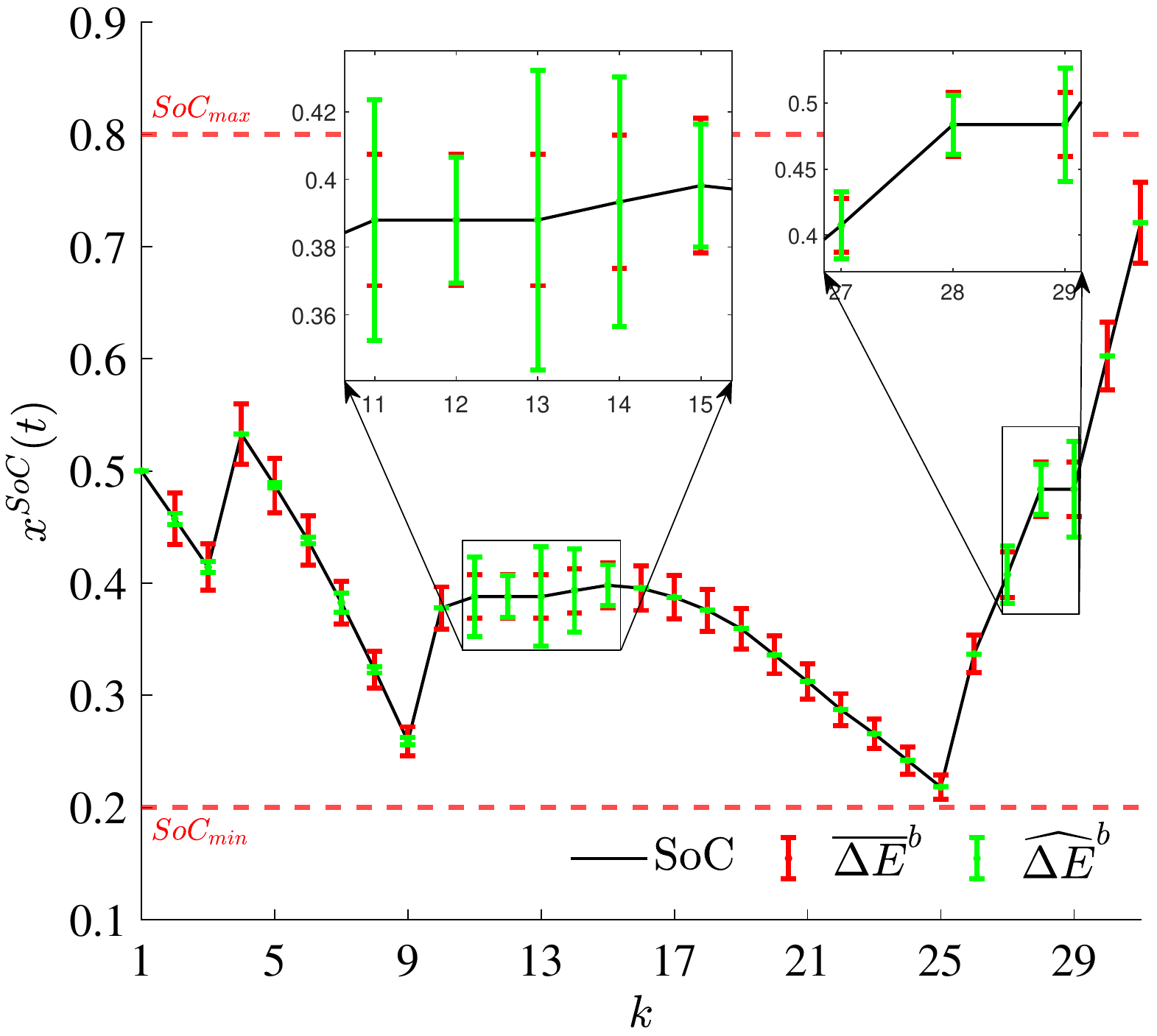}%
    \label{fig:soc_noEB}}
    \\
    \subfloat[With bounds on the energy deviation.]{\includegraphics[width=\columnwidth]{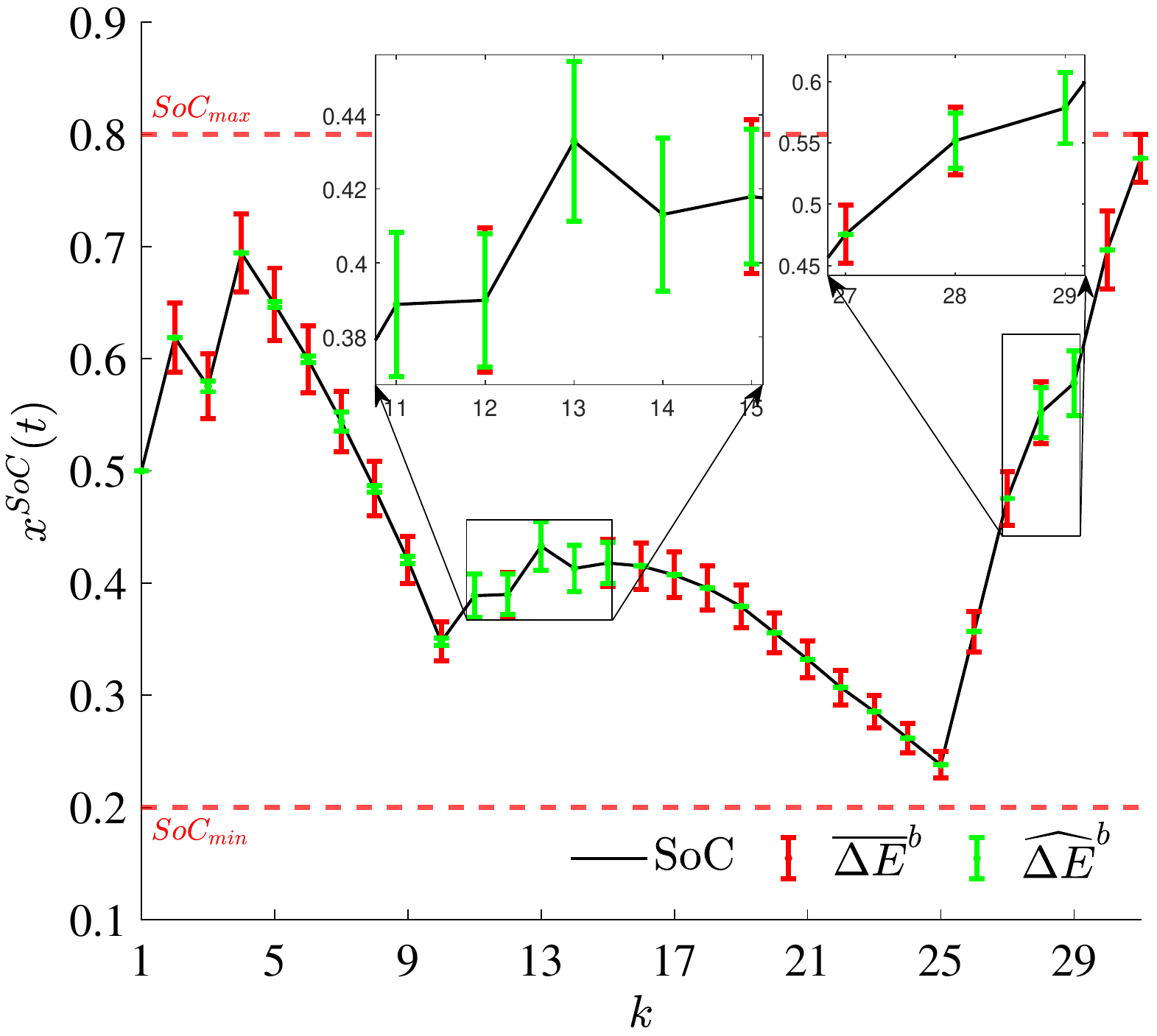}%
    \label{fig:soc_EB}}
    \caption{State of charge trajectories from the optimal scheduling and primary control design of the \gls{ess}, during the simulation period, for the cases of not including bounds on the resulting energy deviation from the participation in \gls{pfc} (top) and the one when including the bounds (bottom).}
    \label{fig:soc_with_bounds}
\end{figure}

An additional comparison demonstrating the additional benefits coming from method \textit{III} over method \textit{II} is illustrated in \cref{fig:soc_with_bounds}.
The resulting $x^{SoC}(t)$ signals from the application of \textit{II} and \textit{III} are depicted in \cref{fig:soc_noEB} and \cref{fig:soc_EB} correspondingly.
Notice that both methods respect the upper and lower \gls{soc} limits, as originally designed in the default method \textit{I}.
Both trajectories seem to follow similar patterns, i.e. initial discharge until $k=9$, smooth re-bouncing and discharge until $k=25$, and charging until the end.
There are however small but important differences, those being clearly depicted in the two zoomed areas ($k=11-15$ and $k=27-29$), where the maximum allowed energy deviation $\overline{\Delta E^{b}}(t)$ from the frequency support offered by the \gls{ess} is illustrated with red error bars and the calculated upper bound $\widehat{\Delta E^{b}}$ is illustrated with green error bars.
It is evident that, for case \textit{II}, there would be requirements for the \gls{ess} that could cause it to violate the upper bound  $\widehat{\Delta E^{b}} > \overline{\Delta E^{b}}(t)$ at $k=11, 14, 27, 29$, while the method in \textit{III} controlled the \gls{ess} in a way that $\widehat{\Delta E^{b}} \le \overline{\Delta E^{b}}(t)$ is guaranteed for the whole period.

\section{Conclusions}
\label{sec:Conclusions}

Achieving optimal energy management in isolated power systems with energy storage and sudden load variations cannot be decoupled from ensuring their secure and robust operation, especially under the presence of intermittent renewable power sources.
Even though decisions related to techno-economical operation are conventionally taken in the discrete time, those will inevitably affect the system's stability in its continuous operation and vice versa.
To address this problem, this article proposed an energy management algorithm capable of integrating both higher time scale economic objectives and lower time scale stability constraints under adaptive uncertainty considerations.
Additional constraints regarding the optimal use of the energy storage for providing flexibility and frequency support with bounded interaction between both services were proposed.
A \gls{milp} formulation was derived and validated through time-domain simulations for an isolated offshore \gls{og} platform integrating wind power.
The results indicated that, under the proposed adaptive uncertainty framework, optimal decisions with dynamic frequency stability guarantees could be achieved and secured the system under an adaptive assessment of possible active power perturbations.
This also reduced the conservatism from setting fixed damping and inertia requirements based on the expected worst-case and allowed better scheduling and operation of the \glspl{gt} for longer periods.
At the same time, the optimal sharing of primary frequency control contribution from conventional generators and the energy storage was found, while ensuring a tolerable impact on the storage optimal state of charge schedule and a negligible impact on the rest energy management objectives.


%

\appendices



\ifCLASSOPTIONcaptionsoff
  \newpage
\fi



\bibliography{IEEEabrv.bib, Biblio/lit_vol_01.bib}
\bibliographystyle{my3IEEEtran}

%









\end{document}